\documentclass[aps,prb,reprint]{revtex4-1}
\usepackage{graphicx}
\usepackage{amsmath}
\usepackage{soul,color}
\sethlcolor{green}
\begin{document}

\title{Negative thermal expansion in CdSe quasi-two-dimensional nanoplatelets}

\author{Alexander I. Lebedev}
\email[]{swan@scon155.phys.msu.ru}
%\thanks{}
\affiliation{Physics Department, Moscow State University, 119991 Moscow, Leninskie gory, Russia}

\date{\today}

\begin{abstract}
The in-plane coefficient of thermal expansion (CTE) for CdSe nanoplatelets with
the zinc-blende structure containing from two to five monolayers is calculated
from first principles within the quasiharmonic approximation. A comparison of
the obtained results with those for bulk CdSe with both the zinc-blende and
wurtzite structures finds a significant increase in the magnitude of negative
CTE and the temperature range of its observation in nanoplatelets. The
main contribution to the negative thermal expansion in CdSe nanoplatelets is
given by the out-of-plane flexural ZA mode and in-plane optical $E$~modes that
arise from the folding of TA phonon of bulk CdSe.

\texttt{DOI: 10.1103/PhysRevB.100.035432}
\end{abstract}

%\pacs{}

\maketitle

\section{Introduction}

The physical properties of nanoplatelets are qualitatively different from those
of bulk materials. In addition to the well-known size effect in semiconductors,
the quasi-2D character of nanoplatelets results in changes of their vibrational
spectra and associated physical properties (sound velocity, heat capacity, thermal
conductivity, infrared absorption, etc.) as well as in more subtle effects
associated with electron-phonon interaction, such as the temperature dependence
of the forbidden band gap. Numerous experiments have shown that the temperature
dependence of the forbidden band gap in nanoscale semiconductors depend on the
size of nanoparticles~\cite{PhysRevLett.81.3539,ChemPhysLett.439.65,JPhysChemC.114.15280}
as well as on their shape.~\cite{ApplPhysLett.90.093104}  Since one of the
contributions to this dependence results from the thermal expansion, the study
of this effect in nanoplatelets is an actual problem.

The nanoparticles of cadmium chalcogenides, and in particular CdSe, have
attracted considerable attention due to their unique optical properties which
can be controlled by the size effect or by creating nanoheterostructures. These
properties are promising for various applications in nano- and optoelectronics
(see Ref.~\onlinecite{PhysRevB.95.165414} and references therein). It is known
that the thermodynamically stable modification of bulk CdSe is the hexagonal
wurtzite structure. However, CdSe can exist in a metastable cubic zinc-blende
(sphalerite) structure. Due to this metastability, the experimental data on the
properties of zinc-blende CdSe are limited. In particular, the thermal expansion
was studied only for the hexagonal CdSe~\cite{JMaterSci.35.2451} (see also
the preliminary data in Refs.~\onlinecite{PhysRev.129.1009, TPRC.13.1185}). At
low temperature, this phase exhibits a negative thermal expansion.~\cite{ReeberRR1968}
CdSe nanoplatelets, depending on the preparation conditions, can be obtained in
both the zinc-blende and wurtzite modifications.~\cite{JAmChemSoc.130.16504,
AngewChem.48.6861}

Theoretical calculations of the thermal expansion in bulk CdSe with
wurtzite~\cite{CommunTheorPhys.50.220} and zinc-blende~\cite{CommunTheorPhys.50.220,
ComputMaterSci.50.1460} structures were performed earlier using the
quasiharmonic Debye model. These calculations, however, can hardly be considered
as reliable since the existence of the optical vibrations was neglected in them,
and the anisotropy of the crystal structure was not taken into account when
calculating the properties of hexagonal CdSe.

In this work, the temperature dependence of the in-plane coefficient of thermal
expansion (CTE) for CdSe nanoplatelets with the zinc-blende structure and a
thickness from two to five monolayers (ML) as well as for bulk CdSe crystals with
both the zinc-blende and wurtzite structures are calculated from first principles
within the quasiharmonic approximation. The influence on the thermal expansion
of the F, Cl, and Br terminating atoms, which are used to compensate for an
extra charge produced by an extra Cd layer on the surface of the nanoplatelets,
was also studied. So far, the thermal expansion of quasi-2D structures was
investigated using this technique for graphene,~\cite{PhysRevB.71.205214,
PhysRevB.89.035422,JPhysCondensMatter.27.083001,RSCAdv.7.22378}
hexagonal BN,~\cite{PhysRevB.89.035422,RSCAdv.7.22378}
transition metal dichalcogenides,~\cite{PhysRevB.90.045409,PhysRevB.89.035422,
ChinPhysB.24.026501} black and blue phosphorene,~\cite{PhysRevB.92.081408,
PhysRevB.94.165433,PhysLettA.380.2098} silicene,~\cite{PhysRevB.94.165433} and
germanene.~\cite{PhysRevB.94.165433}
Molecular dynamics is another technique that was used to study the thermal
expansion in quasi-2D carbon nanostructures~\cite{PhysRevB.68.035425,
PhysRevLett.102.046808,PhysRevLett.106.135501} and
BN.~\cite{JPhysCondensMatter.25.045009}

\section{Quasiharmonic approximation}

The thermal vibrations of atoms in solids are not strictly harmonic. The
nonlinear dependence of interatomic forces on the interatomic distances
results in the appearance of anharmonicity, the interaction of different
vibrational modes, and the thermal expansion of a solid. Rigorous treatment
of the anharmonicity requires the use of molecular dynamics with quantum-mechanical
calculation of forces acting on atoms, but often a fairly good estimate of
the thermodynamic properties can be obtained using the quasiharmonic
approximation (QHA).

The QHA assumes that, when the temperature is varied,
the individual vibrational modes remain independent and harmonic, and the
anharmonicity effects can be taken into account via the dependence of phonon
frequencies $\omega_j$ on the unit cell volume~$V_0$. As in ordinary
thermodynamics, in this approach all the details of the microscopic interactions
between atoms are hidden and only the consequences of these interactions
(thermal expansion, changes in vibrational frequencies, etc.) are considered
by expressing them using thermodynamic parameters. This enables one to predict
the macroscopic properties of solids at the thermodynamic level, using such
concepts as temperature~$T$, pressure~$P$, free energy~$F$, volume~$V$, etc.,
but remaining strictly based on first-principles calculations of phonon
frequencies.

In the QHA, the free energy of a crystal unit cell is a sum of
its total energy $E_{\rm tot}$ calculated using the density functional theory and
the free energy of a system of noninteracting harmonic oscillators $F_{\rm vib}$:
    \begin{equation}
    F(V_0,T) = E_{\rm tot}(V_0) + F_{\rm vib}(V_0,T),
    \label{eq1}
    \end{equation}
    \begin{equation}
    F_{\rm vib}(V_0,T) = \frac{1}{N_q} \sum_{j{\bf q}} \bigg[ \frac{\hbar\omega_{j{\bf q}}}{2} + kT \ln \bigg(1 - e^{-\hbar\omega_{j{\bf q}}/kT} \bigg) \bigg].
    \end{equation}
Here the sum runs over all phonon branches~$j$ and all wave vectors ${\bf q}$
of the Brillouin zone; $N_q$ is the number of different wave vectors.

At $T \ne 0$, the thermodynamic equilibrium is reached at a volume~$V_0(T)$
satisfying the condition $\partial F(V_0,T) / \partial V_0 = 0$. Taking into
account Eq.~(\ref {eq1}), this condition can be rewritten as
    \begin{equation}
    \begin{split}
    \frac{\partial F(V_0,T)}{\partial V_0} &= \frac{dE_{\rm tot}(V_0)}{dV_0} + \frac{\partial F_{\rm vib}(V_0,T)}{\partial V_0} \\
    &= - P(V_0) - P_{\rm vib}(V_0,T) = 0, \\
    \end{split}
    \end{equation}
i.e. the thermodynamic equilibrium is reached at
$P(V_0) = \partial F_{\rm vib}(V_0,T) / \partial V_0$. This enables one to calculate the
$V_0(T)$ dependence and find the CTE.

In anisotropic crystals whose symmetry is lower than the cubic one, the volume
and derivatives with respect to volume are not correct thermodynamic parameters.
In this case, an approach can be used in which independent lattice parameters are
used instead of the volume (see, for example, Ref.~\onlinecite{ApplPhysLett.88.061902}).
In this work, the strain tensor $u_{ij}$ and the stress tensor
$\sigma_{ij} = - (1/V_0) \partial F_{\rm vib} / \partial u_{ij}$ are used as parameters
describing the deformation effects.

\section{Calculation details}

The calculations presented in this work were performed within the plane-wave
density functional theory using the \texttt{ABINIT} software package. The local
density approximation (LDA) and optimized norm-conserving separable
pseudopotentials constructed using the RRKJ scheme~\cite{PhysRevB.41.1227}
were used in the calculations.%
    \footnote{For quasi-2D systems, the LDA approximation was shown to provide
    better results as compared to the GGA one.~\cite{PhysRevLett.96.136404}
    Earlier, we used LDA when studying ferroelectric and piezoelectric
    properties of SnS mono- and multilayers.~\cite{JApplPhys.124.164302}}
The cutoff energy was 30~Ha (816~eV); the
integration over the Brillouin zone was carried out using 8$\times$8$\times$8
and 8$\times$8$\times$6 Monkhorst--Pack meshes for cubic and hexagonal
crystals, respectively. When modeling nanoplatelets, the 8$\times$8$\times$1
mesh was used. The relaxation of the unit cell
parameters and atomic positions was carried out until the forces acting on
the atoms became less than $2 \cdot 10^{-6}$~Ha/Bohr (0.1~meV/{\AA}). The
accuracy of calculating the total energy was better than 10$^{-10}$~Ha.

\begin{figure}
    \centering
    \includegraphics{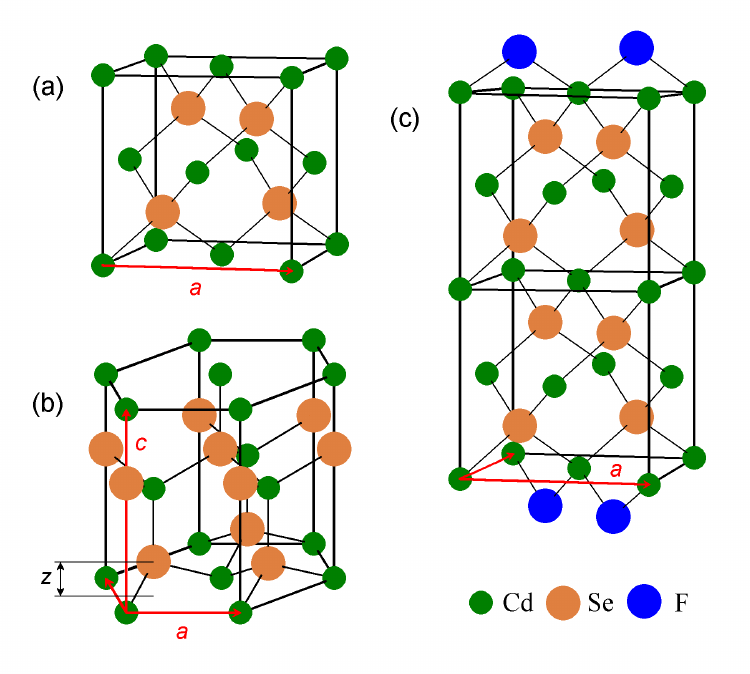}
    \caption{\label{fig1}Structure of bulk CdSe with (a) zinc-blende and (b)
    wurtzite structure and (c) the structure of CdSe nanoplatelet with a thickness
    of 4~ML.}
    \end{figure}

To calculate the volume dependence of the electronic contribution $E_{\rm tot}(V)$
for bulk zinc-blende CdSe [Fig.~\ref{fig1}(a)], we first determined the
equilibrium lattice parameter $a_0$ and the dependence of a mechanical stress
$\sigma_{xx} = \sigma_{yy} = \sigma_{zz}$ in the unit cell as a function of its
isotropic strain $u_{xx} = u_{yy} = u_{zz} = (a - a_0)/a_0$, which was varied
from $-$0.01 to 0.01 in steps of 0.005. When calculating the stress in the
unit cell, we used the ability of the \texttt{ABINIT} program to calculate the
stress tensor using the density functional perturbation theory.~\cite{PhysRevB.32.3792}
The obtained data were then used to calculate the coefficients in the quadratic
approximation $u_{xx} \approx c_1 \sigma_{xx} + c_2 \sigma_{xx}^2$. After that,
for $u_{xx} = -0.01$, 0, and 0.01, the exact values of phonon frequencies were
calculated on the 4$\times$4$\times$4 mesh of wave vectors (64 ${\bf q}$ values,
eight irreducible points in the Brillouin zone). Using the \texttt{anaddb} program,
the phonon frequencies were interpolated on the 64$\times$64$\times$64 mesh,
and the vibrational contribution to the free energy $F_{\rm vib}(u_{xx}; T)$ was
calculated in the temperature range of $T = {}$5--1000~K in steps of 5~K. For
each temperature, the obtained values of $F_{\rm vib}(u_{xx}; T)$ were approximated
by a parabola, the coefficients of its derivative
[$\sigma_{\rm vib} \equiv -(1/V_0) \partial F_{\rm vib} / \partial u_{xx} \approx A + Bu_{xx}$]
were calculated, and finally the $u_{xx}$ values, which are solutions of a system
of two nonlinear equations,
$\sigma_{xx} = -A - Bu_ {xx}$ and $u_{xx} = c_1 \sigma_{xx} + c_2 \sigma_{xx}^2$,
were determined. We note that the calculated value of
$\partial F_{\rm vib} / \partial u_{xx}$ is three times larger than the true value
because the strain was applied three times ($u_{xx} = u_{yy} = u_{zz}$) when
calculating $F_{\rm vib}$. The derivative $du_{xx}/dT$ is the coefficient of linear
thermal expansion $\alpha(T)$ of cubic CdSe.

For bulk CdSe with the wurtzite structure [Fig.~\ref{fig1}(b)], the
calculation scheme was similar. After the equilibrium lattice parameters
($a_0$,\,$c_0$) were found, the internal
stresses $\sigma_{xx}$ and $\sigma_{zz}$ were calculated for 13~pairs of
($u_{xx}$,\,$u_{zz}$) values, namely (0;\,0), ($\pm 0.01$;\,0), (0;\,$\pm 0.01$),
($\pm 0.005$;\,$\pm 0.005$), and ($\pm 0.0025$;\,$\pm 0.0025$). For each strain,
the value of the $z$~parameter describing the relative shift of two hexagonal
sublattices in the wurtzite structure was carefully optimized. The obtained data
were then approximated by formulas
    \begin{equation}
    \begin{split}
    u_{xx} &\approx c_1 \sigma_{xx} + c_2 \sigma_{xx}^2 + c_3 \sigma_{zz} + c_4 \sigma_{zz}^2 + c_5 \sigma_{xx}\sigma_{zz}, \\
    u_{zz} &\approx d_1 \sigma_{xx} + d_2 \sigma_{xx}^2 + d_3 \sigma_{zz} + d_4 \sigma_{zz}^2 + d_5 \sigma_{xx}\sigma_{zz}. \\
    \end{split}
    \label{eq4}
    \end{equation}
The free energy $F_{\rm vib}(u_{xx},u_{zz};T)$ was calculated for the same
($u_{xx}$,\,$u_{zz}$) pairs in the temperature range of $T = {}$5--1000~K in
steps of 5~K. For each temperature, by approximating the obtained values of
$F_{\rm vib}$ by a quadratic form
$F_{\rm vib} \approx e_0 + e_1 u_{xx} + e_2 u_{xx}^2 + e_3 u_{zz} + e_4 u_{zz}^2 + e_5 u_{xx}u_{zz}$,
the derivatives $\sigma_{xx} = -(1/V_0) \partial F_{\rm vib} / \partial u_{xx}$ and
$\sigma_{zz} = -(1/V_0) \partial F_{\rm vib} / \partial u_{zz}$ were calculated. After
that, taking into account Eq.~(\ref{eq4}), a system of nonlinear equations was
solved and the values of $u_{xx}$ and $u_{zz}$ for each $T$ were calculated.
Their derivatives with respect to temperature are the coefficients of linear
thermal expansion $\alpha_{xx}(T)$ and $\alpha_{zz}(T)$ of hexagonal CdSe.

CdSe nanoplatelets studied in this work were the [001]-oriented zinc-blende
platelets with a thickness from two to five monolayers, both surfaces of
which were terminated with cadmium atoms [Fig.~\ref{fig1}(c)].
In order to compensate for an extra charge produced by an extra Cd layer,
the simplest way of charge compensation using F, Cl, or Br terminating atoms
was considered. The modeling of the nanoplatelets was carried out on supercells
to which a vacuum gap of 20~{\AA} was added to consider nanoplatelets as
noninteracting. The symmetry of supercells is described by the tetragonal
$P{\bar 4}m2$ space group. In Ref.~\onlinecite{PhysRevB.95.165414} it was shown
that the most energetically favorable position of terminating atoms is the
bridge position, in which the atoms enter the positions of missing Se atoms.

For these nanoplatelets, the $\sigma_{xx}$($u_{xx}$) curves and phonon spectra
were calculated for biaxial strains $u_{xx} = u_{yy}$ equal to 0, $\pm$0.01,
and $\pm$0.02, with full relaxation of all atomic positions for each strain.
The exact phonon frequencies calculated on the 8$\times$8$\times$1 mesh of
wave vectors ${\bf q}$ were then used to calculate the free energy
$F_{\rm vib}(u_{xx})$ on the 64$\times$64$\times$8 mesh of interpolated frequencies.

\section{Thermal expansion of bulk CdSe}

\begin{figure}
    \centering
    \includegraphics{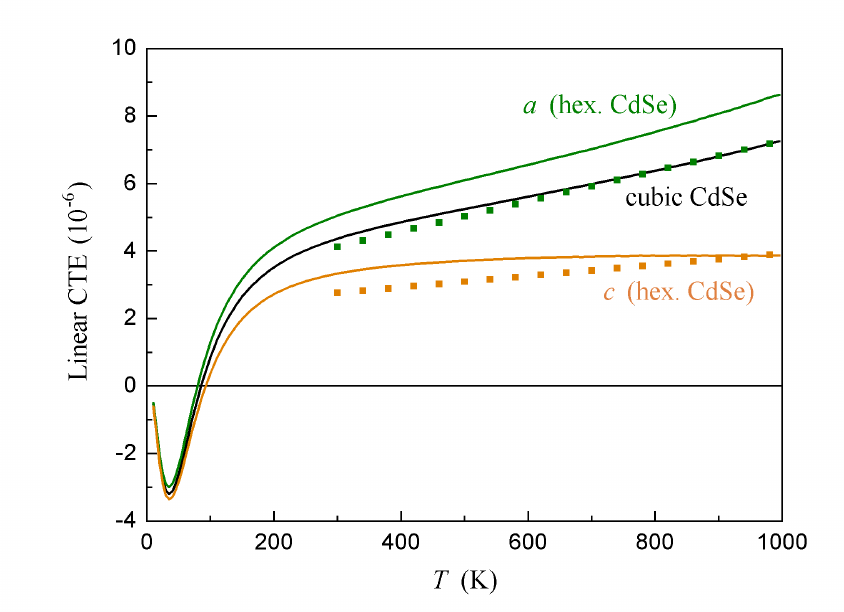}
    \caption{\label{fig2}Coefficient of linear thermal expansion for bulk CdSe
    crystals with zinc-blende and wurtzite structures. The points are experimental
    data for hexagonal CdSe.~\cite{JMaterSci.35.2451}}
    \end{figure}

The calculated temperature dependence of CTE for bulk CdSe with zinc-blende and
wurtzite structures as well as the available experimental data for hexagonal
CdSe~\cite{JMaterSci.35.2451} are shown in Fig.~\ref{fig2}. It is seen that the
calculated curves are in reasonable agreement with the experiment. A discrepancy
in the behavior of the curves should not be considered negatively as an ideal
agreement of the curves can hardly be expected for the used approximation.
A higher CTE along the $a$~axis for hexagonal CdSe is characteristic of other
semiconductors with the wurtzite structure.~\cite{JMaterSci.35.2451}
The CTE curve for the cubic CdSe is located between two curves for the hexagonal
structure.

An interesting feature of the obtained curves is the negative thermal expansion
at low temperatures. This effect is well known in semiconductors with diamond,
zinc-blende, and wurtzite structures.~\cite{PhysRev.112.136,PhysRev.163.779,
TransAIME.233.698,PhysStatusSolidiB.33.257}  By differentiating Eq.~(\ref{eq1})
first with respect to volume and then with respect to temperature, the linear
CTE of a crystal can be written as
    \begin{equation}
    \begin{split}
    \alpha =& \frac{A_0}{V_0} \frac{1}{N_q} \sum_{j{\bf q}} \gamma_{j{\bf q}} \, \hbar\omega_{j{\bf q}} \frac{d}{dT} \bigg( \frac{1}{e^{\hbar\omega_{j{\bf q}}/kT} - 1} \bigg) \\
    =& \frac{A_0}{V_0} \frac{k}{N_q} \sum_{j{\bf q}} \gamma_{j{\bf q}} \, \frac{(\hbar\omega_{j{\bf q}}/kT)^2}{2[\cosh(\hbar\omega_{j{\bf q}}/kT) - 1]}, \\
    \label{eq5}
    \end{split}
    \end{equation}
where $\gamma_{j{\bf q}} = -d \ln \omega_{j{\bf q}} / d \ln V_0$ is the
Gr{\"u}neisen parameter for mode~$j$ with the wave vector~${\bf q}$ and
$A_0 = 1/(3 B_0)$, where
$B_0$ is the bulk modulus. From this formula it follows that for a negative
thermal expansion to appear, it is necessary that some modes have a negative
sign of~$\gamma$. For most modes in crystals, the interatomic forces weaken
when the lattice is stretched, and therefore the Gr{\"u}neisen parameters
for them are positive. An analysis of the volume dependence of the frequencies
of different modes in zinc-blende semiconductors found negative~$\gamma$
parameters for transverse acoustic (TA) phonons near the $X$ and $L$~points
of the Brillouin zone.~\cite{PhysRevB.43.5024}  This effect was explained by
the influence of the strain on the magnitude of restoring forces that act
on the atoms oscillating in a direction perpendicular to the direction of the
chemical bond. The mechanism of the negative thermal expansion in wurtzite
semiconductors is similar to that proposed for zinc-blende semiconductors;
however, the negative Gr{\"u}neisen parameter in the wurtzite structure may
also appear for TO modes and even LA mode.~\cite{JApplPhys.114.063508}

\section{Phonon spectra and thermal expansion of CdSe nanoplatelets}

\begin{figure}
    \centering
    \includegraphics{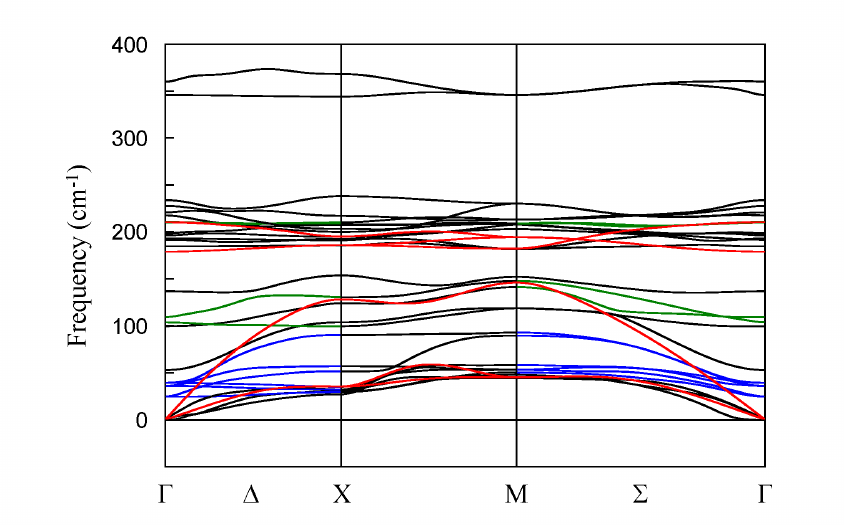}
    \caption{\label{fig3}Phonon spectra of 3~ML CdSe nanoplatelet terminated
    with F~atoms (black lines) and bulk CdSe (red lines). Blue lines show
    acoustic-like $E$~modes with a negative Gr{\"u}neisen parameter and green
    lines show surface modes with a negative Gr{\"u}neisen parameter.}
    \end{figure}

The phonon spectrum of a typical quasi-two-dimensional F-terminated CdSe
nanoplatelet with the zinc-blende structure is shown in Fig.~\ref{fig3}.~\cite{PhysRevB.96.184306}
The two-dimensional character of the phonon spectrum is confirmed by the absence
of a dispersion of the phonon modes in the $q_z$ direction normal to the plane
of the nanoplatelet.

A detailed analysis of phonon spectra of CdSe nanoplatelets was carried out in
Ref.~\onlinecite{PhysRevB.96.184306}. It was shown that the phonon spectrum of a
CdSe nanoplatelet with a thickness of $n$~ML consists of three types of optical
modes: $n$~symmetric quasi-Lamb $A_1$~modes, $n$~antisymmetric quasi-Lamb
$B_2$~modes (in both modes the atomic displacements are out-of-plane), and
$2n$~modes of symmetry~$E$ with the in-plane atomic displacements. Almost all
optical modes have a mixed acoustic+optic character. The lower-energy part
of these modes has an acoustic-like displacement pattern, whereas the
higher-energy part has an optic-like one. The acoustic vibrations are presented
by one LA and one TA phonons, in which the atoms move in the plane of the
nanoplatelet, and the flexural $B_2$~mode (ZA~mode), in which the atoms move
in the out-of-plane direction. Terminating atoms produce six surface modes.
The specific feature of the phonon spectrum of nanoplatelets is a large number
of modes that arise from the folding of acoustic and optical modes of bulk CdSe,
as was demonstrated by the mode projection analysis.

    \begin{figure}
    \centering
    \includegraphics{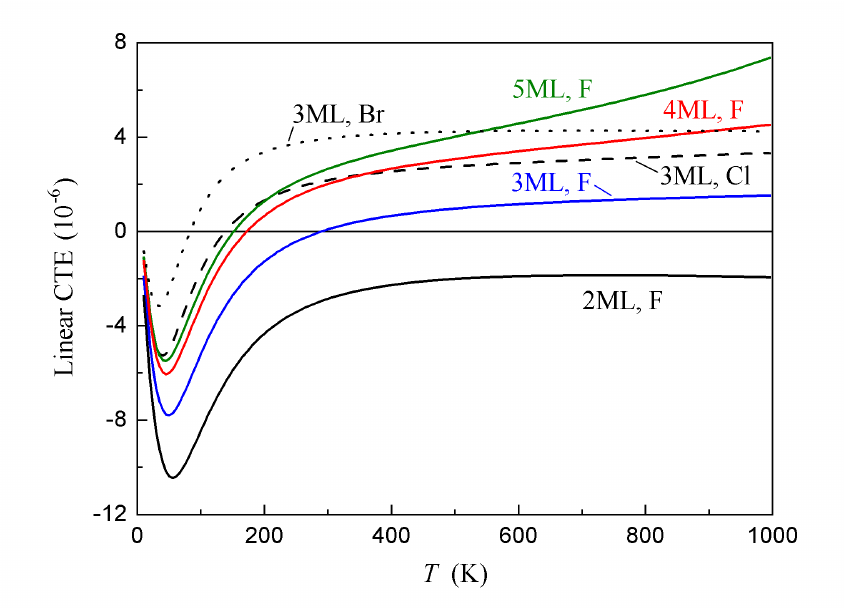}
    \caption{\label{fig4}Coefficient of linear thermal expansion of F-terminated
    CdSe nanoplatelets with a thickness from 2~ML to 5~ML and for 3~ML nanoplatelets
    terminated with F, Cl, and Br atoms.}
    \end{figure}

Calculations of the thermal expansion of nanoplatelets reveal the appearance of
a large negative in-plane CTE which appreciably exceeds its magnitude in bulk
CdSe (compare Figs.~\ref{fig2} and \ref{fig4}). The reasons for this may be an
increase in the negative values of $\gamma$ for some modes and an increase in
the number of modes with negative~$\gamma$ (a change in the phonon
density of states).

    \begin{figure}
    \centering
    \includegraphics{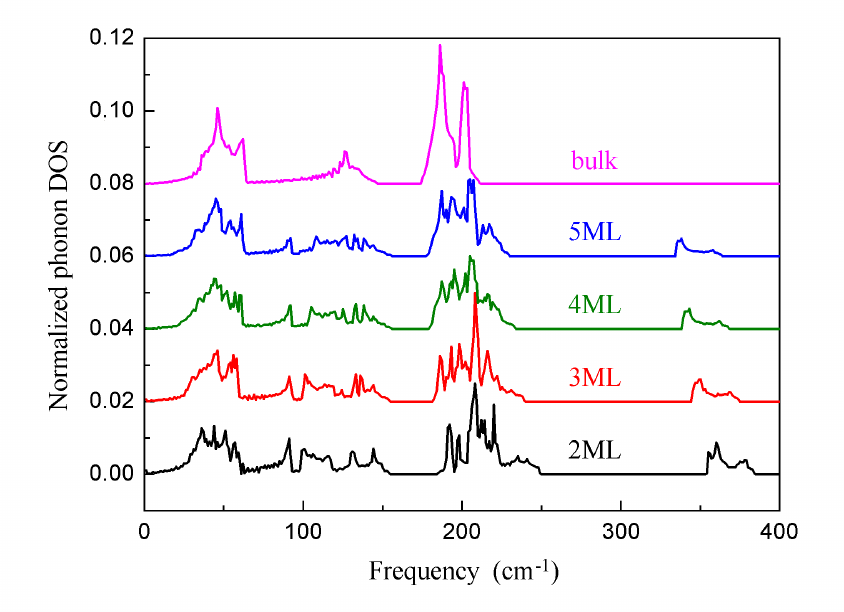}
    \caption{\label{fig5}Normalized phonon density of states for F-terminated
    CdSe nanoplatelets with a thickness from 2~ML to 5~ML and for bulk CdSe.}
    \end{figure}

We consider first a possible change in the phonon density of states (DOS). The
phonon DOS calculated for CdSe nanoplatelets and bulk CdSe are shown in
Fig.~\ref{fig5}. Four regions with frequency ranges of 0--70, 70--160, 160--260,
and 260--400~cm$^{-1}$ are clearly seen in the figure. They correspond to regions
with a predominant contribution of acoustic (ZA, TA, LA) phonons, acoustic-like
$E$ and quasi-Lamb modes, optic-like $E$ and quasi-Lamb modes,
and vibrations of terminating atoms, respectively. An analysis shows that in
going from bulk CdSe to the thinnest of nanoplatelets, the number of
modes in the 0--70 and 160--260~cm$^{-1}$ regions systematically decreases,
whereas the number of phonon modes in the 70--160~cm$^{-1}$ region
increases by more than 1.5~times. One can see, however, that the changes in the
phonon DOS are insufficient to explain the strong increase in the negative
in-plane CTE in nanoplatelets.

    \begin{figure}
    \centering
    \includegraphics{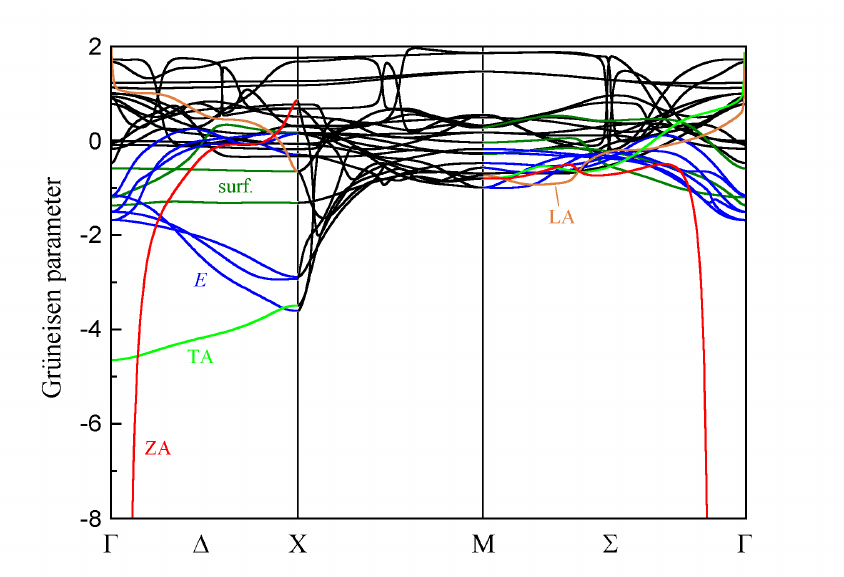}
    \caption{\label{fig6}Gr{\"u}neisen parameter of different modes in
    F-terminated CdSe nanoplatelet with a thickness of 3~ML as a function of
    wave vector.}
    \end{figure}

Calculations of the Gr{\"u}neisen parameters $\gamma_j = -d \ln \omega_j / d \ln a$
($a$ is the in-plane lattice parameter of the nanoplatelet) find a large number
of modes with negative~$\gamma$ (Fig.~\ref{fig6}). In contrast to bulk CdSe, in
which the only mode that has a negative~$\gamma$ is the TA mode (see Fig.~S1
in the Appendix), in nanoplatelets, in addition to ZA
and TA acoustic modes, there are $n$~optical $E$~modes ($n$ is the
thickness of the nanoplatelet) and two surface $E$~modes which have a
negative~$\gamma$. All quasi-Lamb modes ($A_1$ and $B_2$) as well as the
highest-energy vibrational mode of terminating halogen atoms have $\gamma> 0$.
As was shown in Ref.~\onlinecite{PhysRevB.96.184306}, the low-frequency optical
$E$~modes arise from the folding of TA phonon mode which is responsible for
a negative thermal expansion in bulk CdSe. From Fig.~\ref{fig6} it is seen that
a region with a large negative~$\gamma$ for the ZA mode spans over a small part
of the Brillouin zone near the $\Gamma$~point, and so its contribution to the
negative CTE is limited. In contrast, the region of smaller negative~$\gamma$
for TA, $E$, and surface modes is located near the $\Delta$~axis and spans over
a larger volume of ${\bf k}$ space. That is why its contribution to the negative
thermal expansion may be comparable with that
of the ZA mode, especially if one takes into account that the number of the
optical modes with negative~$\gamma$ exceeds the number of acoustic modes.

When the terminating F~atoms are replaced by Cl and then by Br, the phonon
DOS remains nearly unchanged (see Fig.~S2 in the Appendix),
but the magnitudes of both positive and negative
values of $\gamma$ decrease. The averaged $\gamma$ value in the frequency range
above 160~cm$^{-1}$ decreases strongly. These changes explain the evolution
of the CTE curves on Fig.~\ref{fig4} when increasing the mass of the
terminating atom.

A comparison of the Gr{\"u}neisen parameters for modes in F-terminated
nanoplatelets with different thickness shows that, when increasing thickness,
the magnitude of the negative $\gamma$ values in two low-frequency regions of
modes decreases strongly, whereas the $\gamma$ values for modes associated with
vibrations of terminating F~atoms increase monotonically. From Eq.~(\ref{eq5})
it follows that, at a given temperature, the CTE value is approximately
proportional to the averaged $\gamma$ value of all modes with an energy of
$\hbar \omega \lesssim 3kT$. This explains all the systematic changes in the
CTE curves observed in Fig.~\ref{fig4}.
A wider temperature range in which the negative thermal expansion is observed
in CdSe nanoplatelets as compared to bulk CdSe (compare Figs.~\ref{fig2}
and \ref{fig4}) is explained by the fact that the energy of TA phonon in bulk CdSe
does not exceed 59~cm$^{-1}$, whereas in the nanoplatelets the energies of
modes with negative~$\gamma$ reach $\sim$140~cm$^{-1}$ (Fig.~\ref{fig3}).

    \begin{table}
    \caption{\label{table1}Elastic compliance moduli of F-terminated
    CdSe nanoplatelets and bulk material (in $10^{-2}$~GPa$^{-1}$).}
    \begin{ruledtabular}
    \begin{tabular}{ccccccc}
    Parameter & 2~ML     & 3~ML     & 4~ML     & 5~ML     & 6~ML     & bulk \\
    \hline
    $S_{11}$  &    5.477 &    4.707 &    4.395 &    4.218 &    4.105 &    3.607 \\
    $S_{12}$  & $-$4.101 & $-$3.057 & $-$2.618 & $-$2.372 & $-$2.209 & $-$1.524 \\
    $S_{66}$  &    3.653 &    3.743 &    3.799 &    3.832 &    3.856 &    4.017 \\
    \end{tabular}
    \end{ruledtabular}
    \end{table}

It is interesting that, in thick nanoplatelets, the CTE values at high temperatures
quickly reach the values for bulk cubic CdSe. In our opinion, the reason for this
is that the mechanical structure of nanoplatelets is not as rigid as the bulk
cubic structure. The calculations (Table~\ref{table1}) show that three components
of the elastic compliance tensor $S_{\mu\nu}$ for nanoplatelets change slowly
with increasing their thickness and tend towards the values for the cubic phase.
However, for 3D and quasi-2D systems the coefficient $A_0$ that enters
Eq.~(\ref{eq5}) should be different: it is $S_{11}+2S_{12}$ for cubic crystals
and $S_{11}+S_{12}$ for nanoplatelets. This difference is a consequence of
the fact that one cannot control the thickness of the quasi-2D system when applying
the in-plane strain. Using the data from Table~\ref{table1}, one can
show that in nanoplatelets the coefficient $A_0$ is indeed larger than
in the bulk crystal and it increases with increasing the nanoplatelet thickness.

Another interesting effect is the negative CTE observed in the 2~ML nanoplatelet
in the whole temperature range. Formally, according to Eq.~(\ref{eq5}), at high
temperature this situation is possible if the sum of Gr{\"u}neisen parameters over
all branches and wave vectors is negative. Direct calculations show that this is
the case. The physical explanation of this effect is simple: the out-of-plane
thermal motion of very thin nanoplatelets results in shrinking of the nanoplatelets
in the basal plane.

First-principles calculations of the negative thermal expansion in monolayer
graphene,~\cite{PhysRevB.71.205214,PhysRevB.89.035422,JPhysCondensMatter.27.083001,
RSCAdv.7.22378}  $h$-BN,~\cite{PhysRevB.89.035422}  transition metal
dichalcogenides,~\cite{PhysRevB.89.035422,ChinPhysB.24.026501}  and blue and black
phosphorene~\cite{PhysRevB.94.165433,PhysLettA.380.2098} have shown that it is
mainly determined by the ZA-phonon---a flexural mode, which has a negative
$\gamma$~value and in which atoms move normal to the monolayer. In monolayer
graphene, there exists an additional optical ZO mode with a small negative~$\gamma$
($\gamma > -1$),~\cite{PhysRevB.95.085435}  but the frequency of this mode is
$\sim$900~cm$^{-1}$ and so its contribution to the CTE can be seen well above
the room temperature. In bilayer and multilayer graphene, there appears a new
optical ZO$^\prime$ mode, in which the out-of-plane atomic displacements are
in phase in the same layer and out of phase in adjacent layers. This mode has
a negative~$\gamma$ ($\gamma > -7$).~\cite{PhysRevB.95.085435}  A similar mode
with a negative~$\gamma$ is typical of graphite.~\cite{PhysRevB.71.205214}
In black phosphorene, the ZO mode with a frequency of $\sim$130~cm$^{-1}$ is
also characterized by a negative $\gamma$.~\cite{PhysLettA.380.2098}
We note that all the above-mentioned optical modes in monolayers are polarized
normal to the nanoplatelet. In MoTe$_2$ and WTe$_2$, in which the monolayer
consists of three atomic layers, in addition to the ZA~mode with a
negative~$\gamma$ there appear acoustic LA and TA modes with an \emph{in-plane}
polarization, which also have small negative $\gamma$~values
($\gamma > -1.4$).~\cite{ChinPhysB.24.026501}
The TA and LA modes with a negative~$\gamma$ were also observed in monolayers
of silicene, germanene,~\cite{PhysRevB.94.165433}  and blue and black
phosphorene.~\cite{PhysLettA.380.2098}

An analysis of Fig.~\ref{fig6} finds a large contribution of the ZA and TA
acoustic modes to the negative thermal expansion in CdSe nanoplatelets. However,
the most interesting result is that the negative~$\gamma$ is also characteristic
of the \emph{in-plane optical and surface} modes of $E$~symmetry. It looks like
this is a first observation of negative~$\gamma$ for optical modes with in-plane
polarization in quasi-2D systems. This result is not much surprising if one
recalls that these optical modes originate from the folding of
the TA phonon mode of bulk CdSe with a negative~$\gamma$. The origin of
such a behavior results from the zinc-blende structure of our nanoplatelets. For
an unstrained nanoplatelet, the restoring force for acoustic-like optical phonons
with the [110] polarization is produced by bending of Cd--Se chemical bonds. However,
when the nanoplatelet is stretched, the strained chemical bonds produce increased
restoring forces that increase the vibrational frequencies. In our opinion, this
effect can be observed in other quasi-2D multilayer systems with strong enough
interaction between the layers. Indeed, our calculations for twelve quasi-2D
systems (see the Appendix) find a negative~$\gamma$
for in-plane optical modes in two-layer graphene, silicene, germanene, blue and
black phosphorene, SiC, BN, SnS, and TiO$_2$.

The calculated CTE values for CdSe nanoplatelets enable one to estimate the
contribution of the thermal expansion to the temperature dependence of their
forbidden band gap $E_g$. Calculations show that the change in the band gap upon
the biaxial stretching is $dE_g/du_{xx} = {}$4.06~eV for 3~ML nanoplatelet and
3.09~eV for 5~ML one. Thus, the contribution of the thermal expansion to the
temperature dependence $dE_g/dT$ does not exceed $+1.2 \cdot 10^{-5}$~eV/K at
300~K. This means that the large negative $dE_g/dT$ values observed in CdSe
nanoparticles result from the electron-phonon interaction.

In Refs.~\onlinecite{PhysRevB.87.214109,PhysRevB.88.104306} it was
suggested that materials that exhibit negative thermal expansion can also
demonstrate pressure-induced softening, i.e., negative $B' =  dB_0/dP$ values,
where $B_0$ is the bulk modulus. Although quasi-two-dimensional CdSe
nanoplatelets studied in this work are not bulk materials, they exhibit a
negative thermal expansion and so it was interesting to check whether they
will exhibit a pressure-induced softening. The calculations showed (see the
Appendix) that $d(C_{11}+C_{12})/d\sigma_{xx}$
is positive in CdSe nanoplatelets and therefore they do not exhibit a
pressure-induced softening. The absence of this effect is probably due to
the structure of the nanoplatelets, in which tetrahedra are tightly linked
together. This structure has no intermediate chains or other elements that
appear in zeolites and Zn(CN)$_2$, in which the pressure-induced softening
occurs.

\section{Conclusions}

In this work, the in-plane coefficient of thermal expansion of CdSe
nanoplatelets with the zinc-blende structure containing from two to five
monolayers is calculated from first principles within the quasiharmonic
approximation. Like in other quasi-2D systems, the negative thermal expansion
in CdSe nanoplatelets is more pronounced and is observed in a wider temperature
range as compared to bulk CdSe. One of the origins of the negative thermal
expansion is the flexural acoustic (ZA) mode, which is a common feature of all
quasi-2D systems. However, in contrast to all earlier studied quasi-2D systems,
in CdSe nanoplatelets there is another valuable contribution to the negative
thermal expansion resulting from acoustic-like optical $E$~phonons and surface
$E$~modes. It is shown that optical modes with the in-plane polarization and
negative Gr{\"u}neisen parameter are also characteristic of most of two-layer
quasi-2D systems.

The obtained results are quite expected. The acoustic-like optical modes
originate from the folding of the TA mode of bulk CdSe when constructing a
multilayer nanoplatelet. As this TA mode has a negative Gr{\"u}neisen parameter
at all ${\bf q}$, it is not surprising that in nanoplatelets both TA and
acoustic-like optical modes also have a negative Gr{\"u}neisen parameter. The
only difference between the acoustic-like and TA modes is that in the TA mode
the atomic displacements are in-phase in all layers of the nanostructure, whereas
in the optical modes there are phase shifts between these displacements. These
modes have comparable frequencies and Gr{\"u}neisen parameters and so their
contribution to the thermal expansion should be comparable. The contribution
of the flexural ZA mode decreases with increasing thickness of the nanoplatelet
because thick nanoplatelets are less prone to buckling. These simple reasonings
explain how the evolution of the phonon spectra when going from 2D to 3D systems
influences the thermal expansion.

\begin{acknowledgments}
This work was supported by the Russian Foundation for Basic Research (Grant No.
17-02-01068).
\end{acknowledgments}

\appendix

\addtocounter{table}{-1}
\renewcommand{\thetable}{S\arabic{table}}
\addtocounter{figure}{-6}
\renewcommand{\thefigure}{S\arabic{figure}}

\section{}

    \begin{table*}
    \caption{\label{tableS1}Optical in-plane modes with negative~$\gamma$ at the
    $\Gamma$~point in two-layer nanoplatelets of different compounds. The values
    in parentheses are the mode frequencies in cm$^{-1}$.}
    \begin{ruledtabular}
    \begin{tabular}{ccccc}
    Compound  & Space         & Buckling          & Adhesion energy & Modes with negative~$\gamma$ \\
              & group         & parameter ({\AA}) & (meV/{\AA}$^2$) & and their frequencies \\
    \hline
    Graphene  & $P{\bar 3}m1$ &  0.0  &   9.4 & $E_g(27), A_{1g}(888), A_{2u}(891)$ \\
    BN        & $P3m1$        &  0.0  &  21.7 & $E(37), A_1(81), A_1(808), A_1(815)$ \\
    SiC       & $P{\bar 3}m1$ &  0.0  &  61.6 & $E_g(249), A_{1g}(276), A_{1g}(575), A_{2u}(591)$ \\
    Silicene  & $P{\bar 3}m1$ &  0.39 &  57.8 & $E_g(34)$ \\
    Germanene\footnotemark[1] & $P{\bar 3}m1$ &  0.63 &  46.0 & $E_g$(24), $A_{1g}$(104), $A_{1g}$(188), $E_g$(281) \\
    blue P    & $P{\bar 3}m1$ &  1.25 &   9.1 & $E_g(39), A_{1g}(50)$ \\
    black P   & $Pbcm$        &  2.15 &  21.3 & all modes \\
    CdSe (3~ML) & $P{\bar 4}m2$ & --- &  --- & $E(25), E(36), E(39), E(103), E(210)$ \\
    SnS       & $P2_1/m$      &  2.89 &  28.8 & all modes \\
    TiO$_2$ (1~ML)\footnotemark[2] & $Pmmn$ & 2.20 & --- & $B_{3g}(102), B_{1u}(218), B_{3u}(222), B_{1g}(233)$ \\
    GaS (1~ML)  & $P{\bar 6}m2$ &  2.45 & ---  & --- \\
    GaSe (1~ML) & $P{\bar 6}m2$ &  2.44 & ---  & --- \\
    \end{tabular}
    \end{ruledtabular}
    \footnotemark[1]{All frequency--strain curves for germanene are strongly nonlinear. \hfill}

    \footnotemark[2]{The lepidocrocite structure. The buckling parameter is taken
    as a distance between two planes filled by Ti atoms. \hfill}
    \end{table*}

\begin{figure}[h]
\includegraphics{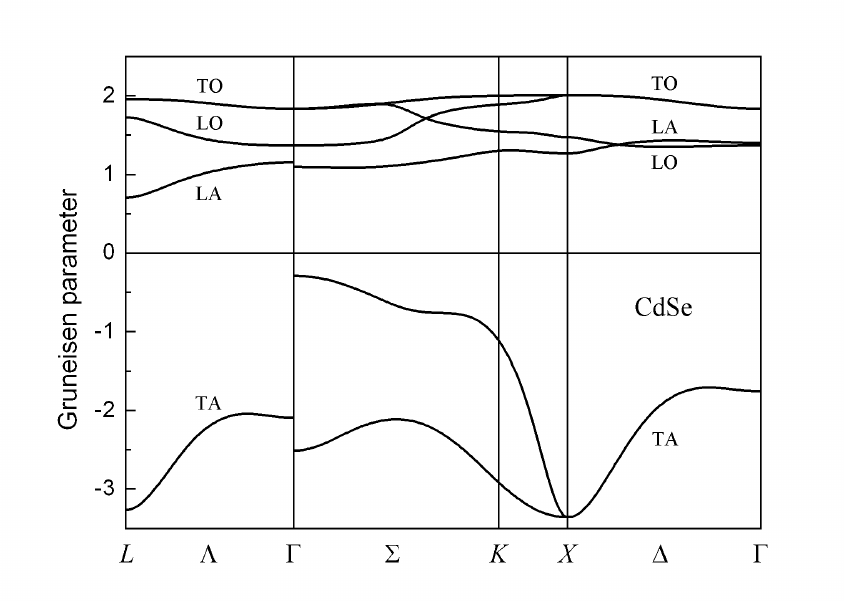}
\caption{\label{figS1}Gr{\"u}neisen parameter for modes in bulk cubic CdSe.}
\end{figure}

An analysis of the results obtained in the main paper has found a large
contribution of the ZA and TA acoustic modes to the negative thermal expansion
in CdSe nanoplatelets. In addition, it was revealed that the negative
Gr{\"u}neisen parameters~$\gamma$ are also characteristic of \emph{in-plane}
optical and surface modes of the $E$~symmetry. This result seems not much
surprising if one reminds that in going from the bulk crystal to nanoplatelets,
the $E$~modes originate from the folding of the TA phonon mode of bulk CdSe (this
branch is characterized by a negative~$\gamma$, see Fig.~\ref{figS1}). The
appearance of negative~$\gamma$ is associated with the zinc-blende structure of
CdSe nanoplatelets. For an
unstrained nanoplatelet, restoring forces for acoustic-like optical phonons with
the [110] displacement pattern are produced by bending of the Cd--Se chemical bonds.
However, when the nanoplatelet is stretched, the strained chemical bonds produce
increased restoring forces that increase the vibrational frequencies. In the main
paper it was supposed that similar in-plane optical modes with a negative~$\gamma$
may occur in other quasi-2D systems which have more than one monolayer and in which
there is a strong interaction between the monolayers.

To check this supposition, we performed calculations of phonon frequencies at
the $\Gamma$~point of the Brillouin zone for twelve strained two-layer quasi-2D
systems and calculated the corresponding Gr{\"u}neisen parameters. The symmetry
of all modes with negative~$\gamma$ and their frequencies are given in
Table~\ref{tableS1}. Additional parameters added to the Table are the buckling
parameter for a single monolayer and the adhesion energy. For monolayers which
contain two levels of atoms and are highly corrugated, the buckling parameter is
taken as a distance between two metal layers. The adhesion energy is taken
as a difference between the total energy of the two-layer nanoplatelet
configuration with the lowest energy and the total energy of two non-interacting
monolayers divided by the contact area.

The calculations reveal that the negative $\gamma$~values are characteristic of
the optical modes with the in-plane polarization in nearly all quasi-2D systems
containing several monolayers (Table~\ref{tableS1}). In graphene, in addition to
the $E$~mode at the $\Gamma$~point ($\gamma = -0.43$), there are two modes at
the $M$~point with the in-plane polarization and negative~$\gamma$. The $E$~modes
with negative~$\gamma$ are observed at the $\Gamma$~point in two-layer boron
nitride ($\gamma = -1.64$), silicon carbide ($\gamma = -3.99$), blue phosphorene
($\gamma = -2.07$), silicene ($\gamma = -9.65$), and germanene ($\gamma = -1.24$).
The modes with the in-plane polarization and negative~$\gamma$ are also observed
at high-symmetry points of the Brillouin zone in all above-mentioned nanostructures.
In black phosphorene and SnS, which have highly
corrugated structures, \emph{all modes} exhibit negative $\gamma$
at the $\Gamma$~point (this means that at least one diagonal component of the
Gr{\"u}neisen tensor is negative). In TiO$_2$, there are four in-plane optical
modes with $\gamma = {}-$(0.89--2.22). It is interesting that in GaS and GaSe
monolayers, which contain two layers of Ga atoms bonded by the metal--metal bond,
the Gr{\"u}neisen parameter for the $E_g$ mode is very small ($|\gamma| < 0.02$).
This means that the frequency of this mode do not depend on the stretching of
monolayers.

We conclude that the negative~$\gamma$ for optical modes with the in-plane
polarization is very typical of quasi-2D systems containing several monolayers.

The buckling parameter and the adhesion energy presented in Table~\ref{tableS1}
are parameters that may characterize the strength of the interlayer interaction.
Nonzero buckling parameter can be regarded as a measure of contribution of hybrids
like $sp^3$ to the chemical bonding: the larger is this parameter, the stronger is
the interlayer interaction. However, this applies only to graphene, silicene,
germanene, BN, SiC, and CdSe, in which the total number of electrons in the outer
shells of a pair of atoms is~8. In phosphorene and SnS, in which the number of
outer electrons is~10, the chemical bonding is based on $p$~orbitals, and the
structure of the monolayer is always highly corrugated. In TiO$_2$, in which the
bonding is highly ionic, the structure consists of edge-sharing TiO$_6$ octahedra.
The adhesion energy can be regarded as a quantitative measure of the interlayer
interaction.

An analysis of the data presented in Table~\ref{tableS1} does not find a clear
correlation between the number of modes with negative~$\gamma$ and a degree of
corrugation of nanoplatelets. Even two-layer graphene and BN bonded with the weakest
van der Waals forces demonstrate the appearance of the in-plane modes with a negative
Gr{\"u}neisen parameter. The only exception are GaS and GaSe, in which the interlayer
bonding has a metal--metal character.

\section{}

As the calculations of the physical properties of quasi-two-dimensional nanoplatelets
are performed on 3D supercells containing a vacuum gap, we need to justify which of
the obtained parameters are proper, i.e. independent of the thickness of this gap.

Both pressures, the stress in the unit cell and $dF_{\rm vib}/da$, are \emph{effective}
parameters: they are normalized by the $c$~lattice parameter of the supercell,
which includes a vacuum gap. However, because they use the same $c$~parameter,
the temperature-induced strain and its derivative with respect to temperature
(the CTE coefficients) are correct. On the other hand, the elastic constants
are effective values because they are calculated as the ratio of the effective
stress to the exact strain. To correctly characterize the elastic properties of
nanostructures, we need to calculate them for the whole nanostructure, by
multiplying the values obtained from the supercell calculation by the $c$~lattice
parameter. The obtained elastic properties, however, are proportional to the thickness
of the nanoplatelet and cannot be compared with the parameters of the bulk material.

In order to get the results, which can be compared with those of bulk material,
we can estimate the elastic properties of nanostructures in the limit of zero
vacuum gap, by using
the actual thickness of the nanostructure corrected for a typical interatomic
distance instead of the $c$~lattice parameter of the supercell. For example, to
calculate elastic properties of CdSe nanoplatelet with a thickness of $n$ ML,
we take its actual thickness (the distance~$d$ between the terminating atoms),
add to it the mean distance between layers, which is equal to $d/2n$, and use
this sum instead of the $c$~parameter. The results given in Table~I of the main
paper were calculated using this method.

\section{}

\begin{table}
\caption{\label{tableS2}In-plane elastic modulus ($C_{11}+C_{22}$) as a function
of the in-plane stress $\sigma_{xx}$ for CdSe 3ML nanoplatelet terminated with
F atoms. All parameters are in GPa.}
\begin{ruledtabular}
\begin{tabular}{cccccc}
$\sigma_{xx}$     & $-$0.1847 & $-$0.0950 &  0.0000 &  0.0978 &  0.2010 \\
$C_{11} + C_{22}$ &   18.191  &   18.708  & 19.260  & 19.773  & 20.304 \\
\end{tabular}
\end{ruledtabular}
\end{table}

The pressure dependence of the in-plane elastic modulus (an analogue of the bulk
modulus~$B$ in 3D systems) for a typical CdSe nanoplatelet is given in Table~\ref{tableS2}.
The differentiation of this dependence gives a dimensionless parameter
$d(C_{11}+C_{22})/d\sigma_{xx}$ (an analogue of $B' = dB_0/dP$) equal to 5.5.
The positive value of this parameter is typical of most solids and shows that
the pressure-induced softening effect (see the main text) is absent in CdSe
nanoplatelets.

\begin{table*}
\caption{\label{tableS3}Frequencies of optical modes at the $\Gamma$~point in 3ML CdSe
nanoplatelet terminated with F, Cl, and Br atoms.}
\begin{ruledtabular}
\begin{tabular}{ccccc}
Mode     & \multicolumn{3}{c}{Frequency (cm$^{-1}$)} & Description\footnotemark[2] \\
\cline{2-4}
symmetry &  F     & Cl    & Br & \\
\hline
$E$   &  25 &    26 &    24 &  acoustic-like \\
$E$   &  37 &    40 &    34 &  acoustic-like \\
$E$   &  40 &    43 &    44 &  acoustic-like \\
$A_1$ &  53 &    50 &    45 &  $s_0$ quasi-Lamb mode \\
$B_2$ & 100 &    95 &    86 &  $a_1$ quasi-Lamb mode \\
$E$   & 104 &    52 &    39 &  surface mode\footnotemark[1] \\
$A_1$ & 137 &   132 &   124 &  $s_1$ quasi-Lamb mode \\
$E$   & 185 &   183 &   167 &  optic-like, becomes a surface mode for Br \\
$E$   & 193 &   190 &   184 &  optic-like\footnotemark[1] \\
$B_2$ & 196 &   190 &   166 &  $a_2$ quasi-Lamb mode \\
$E$   & 210 &   197 &   191 &  optic-like, a surface mode for F\footnotemark[1] \\
$A_1$ & 211 &   202 &   174 &  $s_2$ quasi-Lamb mode \\
$B_2$ & 221 &   214 &   199 &  $a_3$ quasi-Lamb mode\footnotemark[1] \\
$B_2$ & 228 &   223 &   213 &  $a_4$ $z$-polarized surface mode for F \\
$A_1$ & 234 &   226 &   210 &  $s_3$ $z$-polarized surface mode \\
$E$   & 346 &   242 &   195 &  surface mode for F and Cl\footnotemark[1] \\
\end{tabular}
\end{ruledtabular}
\footnotetext[1]{The eigenvector exhibits a strong change for Br termination.}
\footnotetext[2]{For notation of quasi-Lamb modes see Phys. Rev. B {\bf 96}, 184306 (2017).}
\end{table*}

\section{}

\begin{figure}
\includegraphics{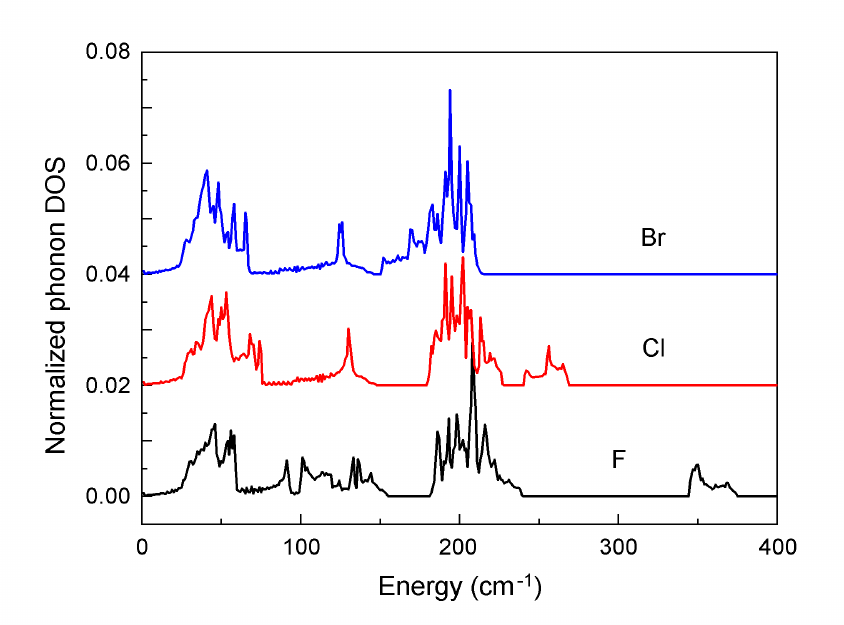}
\caption{\label{figS2}Normalized phonon DOS for 3ML CdSe nanoplatelets terminated
with F, Cl, and Br. A strong variation in the position the high-energy surface
mode results from a strong increase in the mass of terminating atom (see also
Table~\ref{tableS3}).}
\end{figure}

The influence of different terminating atoms on the phonon frequencies at the
$\Gamma$~point and on the phonon density of states for CdSe nanoplatelet with a
thickness of 3ML are presented in Fig.~\ref{figS2} and Table~\ref{tableS3}.

\vspace{\fill}

\newpage

%\bibliography{all}

\providecommand{\BIBYu}{Yu}

\end{document}